\documentclass [a4paper, 12pt]{article}
\usepackage{psfrag}
\usepackage{epsfig}
\usepackage{graphicx}
\usepackage{epstopdf}
\begin{document}

\date{}
\author{Nababrata Ghoshal,
Soumyajit Pramanick$^\#$,
Sudeshna DasGupta$^\#$\\
~and 
Soumen Kumar Roy
\footnote{Corresponding author. E-mail: roy.soumenkumar@gmail.com
}\\ 
Department of Physics, Mahishadal Raj College,\\
Mahishadal, Purba Medinipur, West Bengal, INDIA\\
$^\#$Department of Physics, Lady Brabourne College,\\
Kolkata 700017, INDIA\\
$^\ast$Department of Physics,\\
Jadavpur University, Kolkata - 700 032, INDIA \\
}
\title {
High accuracy Monte Carlo study  of dispersion model of biaxial liquid crystals
 } 
\maketitle

\begin{abstract}
We present a high accuracy Monte Carlo simulation study of the Isotropic - Nematic phase transition of a lattice dispersion model of biaxial liquid crystals. 
The NI coexistence curve terminating at the Landau critical point have been determined using multiple histogram reweighting technique.  
 A close investigation reveals  
a sharp departure in the nature of the $N$-$I$ coexistence curve in  temperature- biaxiality parameter phase diagram in comparison
to the earlier theoretical (either mean-field or computer simulation) predictions.
The coexitence curve shows a change in curvature  with increasing value of the degree molecular biaxiality.  
 
\end{abstract}
\section{INTRODUCTION}

\indent  In recent years, a great deal of attention has been devoted to investigations of
 the phase transformations in the thermotropic liquid crystals composed of bent-core molecules \cite{sluc, nio}. 
Such molecules can be assumed to possess $D_{2h}$ symmetry and are commonly reffered as biaxial molecules in contrast to 
the conventional uniaxial nematogenic molecules having $D_{\infty h}$ symmetry. It is well known from Landau-deGennes (LDG) \cite{degen} and Maier-Saupe mean field theories that the isotropic to nematic phase transition in thermotropic liquid 
crystals  composed of cylindrically symmetric molecules is weakly first order. This has been confirmed by  
experiments \cite{stin, gram} as well as by computer simulatons \cite{zhang, fabbri}.
In a more recent experimental study, Wiant $et~al.$ \cite{wiant} observed that the isotropic ($I$) to the 
uniaxial nematic ($N_U$) transition for LCs composed of 
biaxial (bent-core) molecules is notably weaker than conventional thermotropic LCs formed from uniaxial molecules. In their study \cite{wiant} they
measured a very low stability limit of the isotropic phase of their bent-core compounds by measuring $T_{NI} - T^*\approx 0.4^oC$, $T^*$ being the super-cooling limit, compared to the typical calamitic (rod-shaped) liquid crystals for which $T_{NI} - T^* \geq 1^oC$.\\
\indent  Bent-core molecules have exhibited other diverse effects such as formation of chairal phases in achairal molecules \cite{link}, indication of possible
biaxial nematic order ($N_B$) in thermotropic LCs \cite{mad, ach} etc. Formation of microscopic clusters of bent-core molecules in isotropic phase and 
high degree of molecular biaxiality of this new class of nematics are the primary factors for these unconventional behaviours  as has been 
stated in Ref. \cite{wiant}. (The biaxial nature of the constituent bent-core molecules is relevant specifically for the present study and this issue
will be addressed later.)

\indent The possible effects of molecular biaxiality on nematic order have been studied theoretically using a number of techniques. 
These include molecular field treatments \cite{fre, alb, str, luc1, boc, rem, son, to}, computer simulation studies of lattice dispersion models
 \cite{luc2, bis, rom, bat} and off-lattice biaxial Gay-Berne model \cite{ber1, ber2}. All these studies predict sequences of phase transitions,
 from $I$ to $N_U$ at a higher temperature and from $N_U$ to $N_B$ at a lower temperature. Also a direct $I$ to $N_B$ transition is predicted at a particular molecular geometry.\\
\indent Apart from the above observations, molecular field studies \cite{rem, to} have shown that the increase of degree of molecular biaxiality influences the  $I$ - $N_U$ transition in a number of ways.
First,  as molecular biaxiality parameter increases the nematic order parameter $S$ at the phase transition becomes smaller and thus the jump in $S$ at the
$I$ - $N_U$ transition decreases. Second, the transition temperature $T_{NI}$ decreases monotonically with increase in $\lambda$. Third, the difference between the $I$ - $N_U$  transition temperature and the orientational spinodal temperature ($T^\star$) decreases monotonically with increasing $\lambda$ and finally, these two temperatures merges as $\lambda$ approaches its critical value $\lambda=\lambda_C=1/\sqrt{6}$. \\
\indent More recently, a Monte Carlo (MC) simulation study \cite{gho} based on a lattice dispersion model has investigated
 the influences of molecular biaxiality on $I$ - $N_U$ transition using multiple histogram reweighting technique \cite{fer} and the relevant part of  the free energy has been generated for two different systems - one composed of uniaxial molecules and the other of biaxial molecules. Although the work reported 
in Ref. \cite{gho} emphasized on the effect of an external field on uniaxial and biaxial molecules, however, from free energy analysis it established (pointed out) an important fact that molecular biaxiality weakens further the weak first order $I$ - $N_U$ transition. The investigations presented in 
Ref. \cite{gho} were limited to a single value of molecular biaxialty parameter and also the aim of the study was different, namely the effects of 
field on nematic order. So far, there is no other simulations which were able to investigate the influence of biaxility on $I$ - $N_U$ transition. 
The limited number of studies in this area is due to the lack of accuracy in conventional simulation technique necessary to explore the influences of deviation
from cylindrical symmetry of nematogenic molecules on pretransitional behaviours in $I$ - $N_U$ transition.\\ 
\indent In this paper we present an MC study using powerful reweighting technique \cite{fer} on a lattice dispersion model to investigate
the influences of molecular biaxiality on $I$ - $N_U$ transition. We have found that after a certain value of the molecular biaxiality parameter, $\lambda$ 
(to be elaborated later), the nematic-isotropic phase transition temperature behaves anomalously.  
We thus report on an unprecedented biaxiality-induced change of curvature of the isotropic-nematic co-existence curve in temperature - biaxiality parameter phase diagram for a widely studied dispersion model of biaxial molecules \cite{luc2, bis}.\\
\indent The plan of this paper is as follows: in Sec. II we discuss the dispersion model; in Sec. III we provide the technical details of the simulations; in Sec. IV we present the results. Conclusions are presented in Sec. V.
\section{THE MODEL}
\indent Here we consider a system of biaxial prolate molecules
possessing $D_{2h}$ symmetry (board-like), whose centres of mass are
associated with a simple-cubic lattice. 
We use the dispersion potential \cite{luc2,bis} between 
two identical neighbouring molecules (say $i$th and $j$th molecules)           
\begin{equation}\label{e1} 
U_{ij}^{disp}=-\epsilon_{ij} \{R_{00}^2(\Omega_{ij})+2\lambda[R_{02}^2(\Omega_{ij})+R_{20}^2(\Omega_{ij})]
+4\lambda^2R_{22}^2(\Omega_{ij})\}.
\end{equation}
Here $\Omega_{ij}=\{\phi_{ij},\theta_{ij},\psi_{ij}\}$ denotes the triplet 
of Euler angles defining the relative orientation of $i^{th}$ 
and $j^{th}$ molecules; we have used the convention used by Rose \cite{rose}
in defining the Euler angles. 
$\epsilon_{ij}$ is the strength parameter which is assumed 
to be a positive constant ($\epsilon$) when the particles $i$ and $j$ are 
nearest neighbours and zero otherwise. $R_{mn}^L$ are combinations of 
symmetry-adapted ($D_{2h}$) Wigner functions  
\begin{equation}\label{e2}
R_{00}^2=\frac{3}{2}\cos^2\theta-\frac{1}{2},
\end{equation}  
\begin{equation}\label{e3}
R_{02}^2=\frac{\sqrt{6}}{4}\sin^2\theta\cos2\psi,
\end{equation}  
\begin{equation}\label{e4}
R_{20}^2=\frac{\sqrt{6}}{4}\sin^2\theta\cos2\phi,
\end{equation}  
\begin{equation}\label{e5}
R_{22}^2=\frac{1}{4}(1+\cos^2\theta)\cos2\phi\cos2\psi-\frac{1}{2}\cos\theta\sin2\phi\sin2\psi.
\end{equation}  
The parameter $\lambda$ is a measure of the molecular biaxiality 
and for the dispersion interactions, it can be 
expressed in terms of the eigenvalues ($\rho_1$, $\rho_2$, $\rho_3$) of the 
polarizability tensor {\boldmath $\rho$} of the biaxial molecule
$\lambda=\sqrt{3/2}(\rho_2-\rho_1)/(2\rho_3-\rho_2-\rho_1)$.
The condition for the maximum biaxiality 
(also known as the self-dual geometry) is 
$\lambda=\lambda_C=1/\sqrt{6}$. 
$\lambda < \lambda_C$ corresponds 
to the case of prolate molecules whereas $\lambda > \lambda_C$ corresponds
to oblate molecules. 
This dispersion model can successfully reproduce both 
the uniaxial and the biaxial
orientational orders and various order-disorder transitions as a function of 
temperature and molecular biaxiality \cite{luc2,bis}.\\
\indent In our simulations we consider a range of values of biaxiality parameter. For $\lambda=0$ the pair potential takes the usual Lebwohl-Lasher (LL) form \cite{leb} for nematic liquid crystals of perfectly uniaxial molecules which has been 
extensively studied by Zhang $\textit{et al.}$ \cite{zhang}. 
 $0<\lambda\le0.40325$  represents  biaxial systems composed of prolate biaxial 
molecules. For the LL model 
there is a single weak first-order $I-N_U$ transition at a 
dimensionless temperature ($T_{NI}=kT_K/\epsilon$, $T_K$ being the 
temperature measured in Kelvin and $k$ the Boltzmann's constant)
$T=1.1232\pm 0.0001$ \cite{zhang, gho} ($T=1.1232\pm 0.0006$ \cite{fab}). From 
the Monte Carlo results, as reported in \cite{bis, ghoshal}, the biaxial model 
($0<\lambda<0.40325$) is found to exhibit a biaxial-uniaxial phase transition at lower 
temperature and a uniaxial-isotropic transition at 
higher temperature ($T\approx 1.1$). The biaxial nematic-uniaxial nematic
transition is known to be second order while the uniaxial nematic-isotropic
transition is known to be first order. We restrict our simulations within a narrow range of temperature around the  $I$ - $N_U$ transition as discussed 
below.\\ 
\section{COMPUTATIONAL DETAILS}
\indent A series of Monte Carlo (MC) simulations using the conventional 
Metropolis algorithm on a periodically repeated simple cubic lattice,  
for the system size $L= 64$ ($N=L^3$) have been performed.
The system size chosen in our simulations is sufficiently large so that finite size corrections are negligible. 
We have used a range of values of molecular biaxiality parameter, $\lambda$ ($0, 0.150, 0.200, 0.250, 0.300, 0.325, 0.350, 0.375,$ and  $0.40825$).
An orientational move was attempted following 
the Barker-Watts method \cite{bar}.   
For a given  value of $\lambda$ the simulation at
the lowest temperature studied was started from the 
perfectly ordered state.
The simulations at the other temperatures for the same $\lambda$ run in 
cascade starting from an equilibrium configuration at a nearby lower 
temperature.\\
\indent In each simulation histograms of energy, $h(E)$, were 
accumulated. For this we divided the continuous energy range 
(from $-3.0L^3$ to $0$) with a sufficiently small bin width ($\Delta E=1.0$).
In our simulations $10^6$ sweeps or MCS (Monte Carlo steps per site)
for the equilibration and $3\times10^6$ MCS for the production 
run were used.
For the lattice size ($L=64$), the total run length is more 
than $10~000$ times the correlation time. The total run was divided 
into several ($100$) blocks by performing independent simulations 
so that we could compute the jackknife errors \cite{new}.\\ 
\indent In order to analyze the orientational order we have calculated the 
second rank order
parameters $\langle R_{mn}^2\rangle$ following the procedure described
by Vieillard-Baron \cite{vie}. 
According to this, a $\bf{Q}$ 
tensor is defined for the molecular axes associated with a 
reference molecule. For an arbitrary unit vector {$\bf{w}$}, 
the elements of the $\bf{Q}$ tensor are defined as 
$Q_{\alpha\beta}(\bf{w}) = \langle(3 w_\alpha w_\beta-\delta_{\alpha\beta})/2\rangle$,
where the average is taken over the configurations and the subscripts $\alpha$ 
and $\beta$ label Cartesian components of $\bf{w}$ 
with respective to an 
arbitrary laboratory frame. By diagonalizing the matrix one obtains nine 
eigenvalues and nine eigenvectors which are then recombined to give 
the four order parameters 
$\langle R_{00}^2\rangle$, $\langle R_{02}^2\rangle$, 
$\langle R_{20}^2\rangle$ and $\langle R_{22}^2\rangle$  with respect to 
the director frame \cite{camp}. 
Out of these four second rank order parameters 
the usual uniaxial order parameter  $\langle R_{00}^2\rangle$ (or, $S$)
which measures the alignment of the longest molecular symmetry axis 
with the primary director ($\bf{n}$),
is involved in our study because 
we have simulated a very short temperature range ($1.110~-~1.125$)
around $T_{NI}$ within which no biaxial phase occurs.
 The full set of order parameters 
are required to describe completely the biaxial nematic phase of a 
system of biaxial molecules.\\
\indent We have calculated the ordering susceptibility $\chi$ 
from fluctuations in the order parameter :
$\chi = \frac{L^3(\langle {R_{00}^2}^2 \rangle-{\langle R_{00}^2 \rangle}^2)}{T}$.
In order to determine the order parameter and the ordering susceptibility 
one needs to generate a two-dimensional histogram of energy and order parameter.
An alternative approach \cite{pec} is to estimate 
the constant-energy averages (corresponding to each energy bin)
of the order parameter and its square 
from the simulation data.
These averages are used to evaluate the 
order parameter and the corresponding
susceptibility as a function of temperature using the reweighting method.\\
\indent We have derived the relevant part of the free-energy-like functions $A(E)$ from the energy distribution functions \cite{lee1, lee2} $P(E)$ 
using the relation $A(E) = -~ ln P(E)$, where the normalized histogram count $P(E) = h(E)/\Sigma_E h(E)$. 

\section{RESULTS}
\begin{figure}[tbh]
\begin{center}
\resizebox{120mm}{!}{\rotatebox{0}{\includegraphics[scale=1.0]{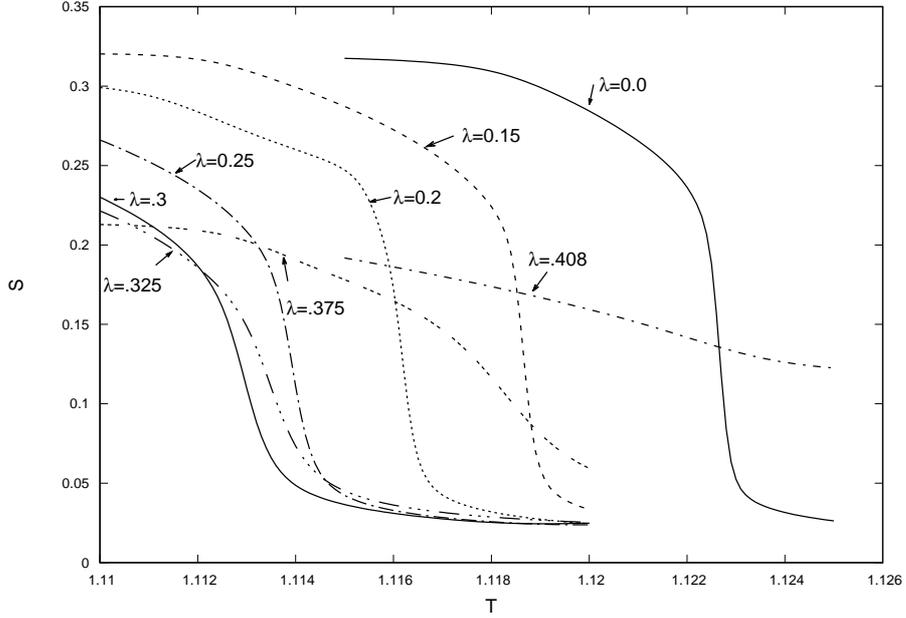}}}
\end{center}
\caption{ Variation of order parameter with reduced temperature for different degrees of molecular biaxiality.}
\label{f1}
\end{figure}

\indent We first present temperature dependence of nematic order parameter (Fig.$\ref{f1}$).
As the degree of the molecular biaxiality increases the jump in the order parameter decreases and the transition shifts towards lower temperature till $\lambda = 0.300$. From $\lambda = 0.325$ the transition shifts towards right. We have studied the system till $\lambda= 1/\sqrt{6}$. The jump at lower biaxialities signifies a first order transition, and the diminishing jump as $\lambda$ gets closer to $1/\sqrt{6}$ implies a weaker first order transition. At $\lambda = 1/\sqrt{6}$ a crossover is observed as expected and the transition is of second order.

\begin{figure}[tbh]
\begin{center}
\resizebox{120mm}{!}{\rotatebox{0}{\includegraphics[scale=0.8]{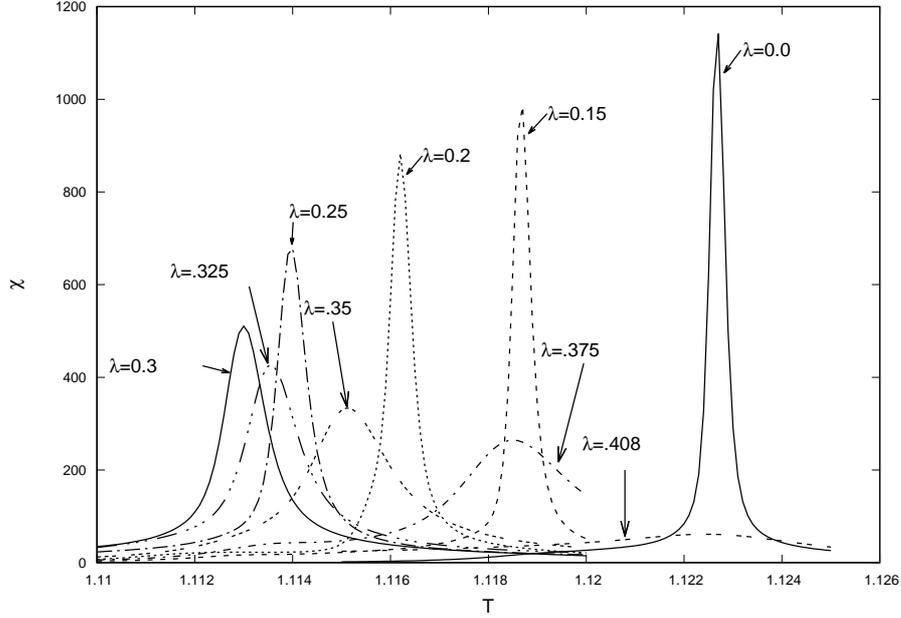}}}
\end{center}
\caption{Variation of  susceptibility with reduced temperature for different degrees of molecular biaxaility. Peak height of the susceptibility curve decreases with increasing biaxiality . Also, the $N-I$ transition temperature decreases with increasing biaxiality until $\lambda= 0.30$ and increases again. }
\label{f2}
\end{figure}

The variation of  susceptibility with reduced temperature for different degrees of molecular biaxaility is given in Fig.$\ref{f2}$. We observe that peak height of the susceptibility curve decreases with increasing biaxiality which shows again the softening of the first-order transition. The temperature at which the $N-I$ transition occurs decreases until $\lambda= 0.3$ and thereafter increases again.

\begin{figure}[tbh]
\begin{center}
\resizebox{120mm}{!}{\rotatebox{0}{\includegraphics[scale=0.8]{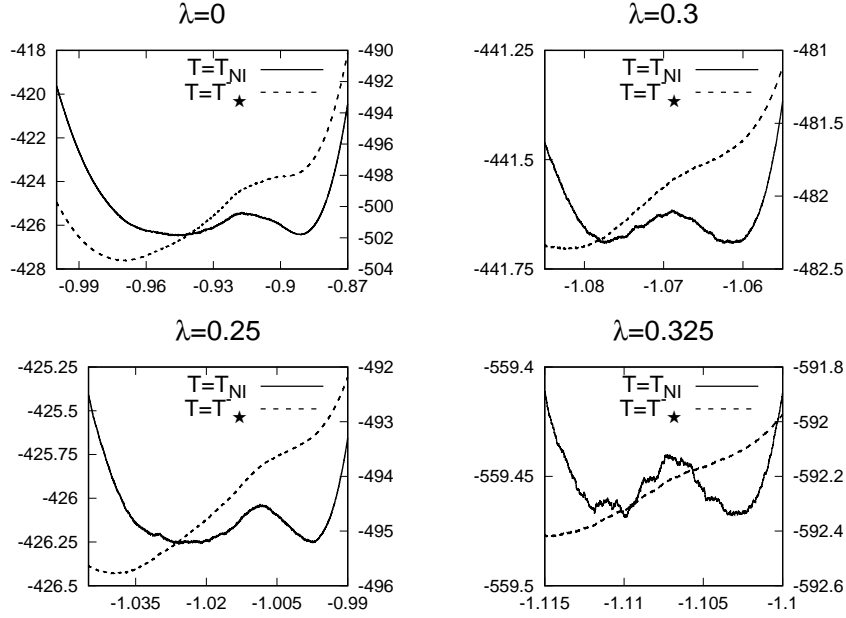}}}
\end{center}
\caption{Free energy vs energy at NI transition for $\lambda=0.15$. }
\label{f3}
\end{figure}

\begin{figure}[tbh]
\begin{center}
\resizebox{120mm}{!}{\rotatebox{0}{\includegraphics[scale=0.8]{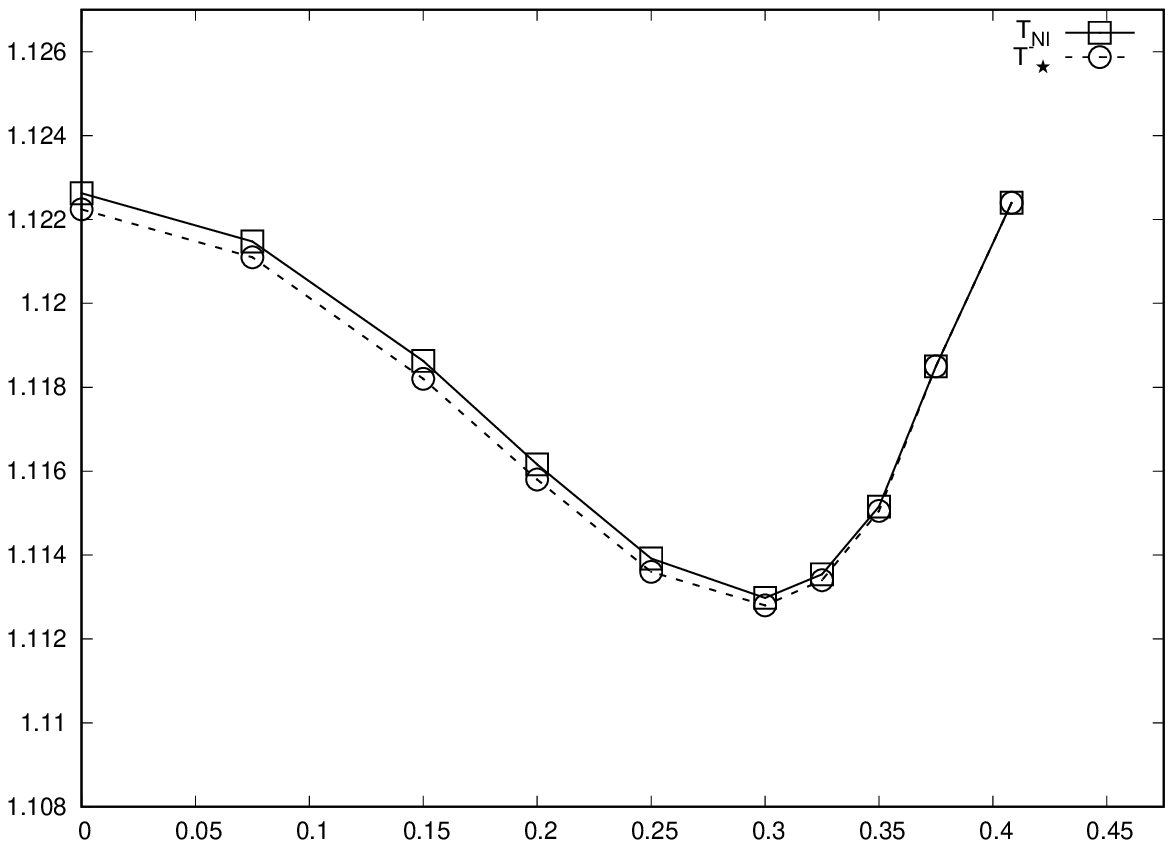}}}
\end{center}
\caption{$T_{NI}$ vs.$\lambda$ Phase diagram  for the nematic isotropic transition. Dependence of transition temperature on degree of biaxiality. The co-existence curve slopes downwards till $\lambda=0.30$ and then slopes upwards till $\lambda=1/\sqrt{6}$.  }
\label{f4}
\end{figure}

We now discuss the pretransitional effects for this model and the influences of molecular biaxiality on these effects. The derived free energy 
functions $A(T,L)$ provide a detailed picture of the stability limits of both the $N_U$ phase and the $I$ phase. Here we shall present 
the stability limit of the $I$ phase only since both the limits are symmetric around the  equilibrium transition for this model. 
The orientational spinodal temperature, $T^*(L)$, for different values of $\lambda$ is estimated as the temperature where the second local minimum of $A(T, L)$ just vanishes as $T$ is gradually lowered below $T_{NI}$. 
The free energy vs energy plots at $N-I$ transition temperature and supercooling temperature for four different values of the biaxiality parameter (Fig.$\ref{f3}$). In each plot two different ordinates have been used to represent the curves corresponding to $T_{NI}$ and $T^{\star}$ on the same plot. The ordinate on the left of each plot corresponds to the free energy vs energy curve at $T_{NI}$ and the ordinate on the right corresponds to the free energy vs energy curve at $T^{\star}$. Figures are plotted column first, with $\lambda$ values $0,0.25,0.30,$ and $0.325$. We can see, from the curves corresponding to $T_{NI}$,  as the value of biaxiality parameter increases, the depth of the free energy well decreases, taking the transition closer to being second order.\\
\indent The change in energy, which corresponds to change of entropy at the transition, also decreases with increasing biaxiality parameter. For $\lambda$ values greater than $0.325$ the depth of the free energy well at transition becomes so small that the structure of the well becomes non-discernible from random fluctuations.\\ 
\indent Finally, in Fig.$\ref{f4}$, we present the coexistence line and the orientational spinodal line in the $\lambda - T$ plane.  We see that both $T_{NI}$
and $T^*$ first decrease with increasing value of the biaxiality parameter and then increase with increasing $\lambda$. Another important 
observation is that the gap between 
these curves decreases monotonically and finally vanishes as $\lambda$ approaches $\lambda_C$. A similar qualitative feature was found in
the molecular-field theory study of To et al \cite{to}. 

\begin{table}[h]
\caption{NI transition temperatures for 
different values of the biaxiality parameter $\lambda$ for the biaxial systems. Estimates of orientational spinodal temperature $T^*$
are also listed for the systems having lower $\lambda$. 
The estimated (jackknife) error in each 
temperature is within $\pm 0.0001$.}
\begin{center}
\begin{tabular}{ c c c c}
\hline
$\lambda$ &  $T_{NI}$(from~$\chi{\it vs}T$) & $T_{NI}$(from~$F(E){\it vs}T$) & $T^*$\\
\hline
 0 &1.1227 &1.1226 & 1.12224 \\
 0.075 &1.1187 &1.1186 & 1.1221 \\
 0.150 &1.1187 &1.1186 & 1.1221 \\
 0.200 &1.1163 &1.1162 & 1.1221 \\
 0.250 &1.1140 &1.1139 & 1.1221 \\
 0.300 &1.1130 &1.1130 & 1.1221 \\
0.325 &1.1136 &1.1135 & 1.1221 \\
0.350 &1.1151 &1.1151 & 1.1221 \\
0.375 &1.1185 &1.1186 & 1.1221 \\
 0.40825 &1.1224 & &  \\
 
 \hline
\end{tabular}
\end{center}
\end{table}
\section{CONCLUSION}
\indent In conclusion we have shown that the NI  coexistence curve of the dispersion model of biaxial molecules is substantially modified. Previous computer simulation studies \cite{luc2, bis, rom, bat} on this model could not explore such behaviour at and around the $I$ - $N_U$ transition. 
This high resolution investigation reveals  
a sharp deviation in the nature of the nematic-isotropic coexistence curve in  temperature- biaxiality parameter phase diagram.
The coexitence curve shows a change in curvature  with increasing value of the degree molecular biaxiality.  A possible reason for the deviation
with increasing value of the degree of molecular biaxiality may be due to the corresponding increase of strength of interactions among the transverse axes of the biaxial molecules.\\
\indent  Other pretransitional phenomena such as, dependence of super cooling temperature on molecular biaxiality etc., have also been observed.\\
\indent Our study, although, is based on a simple lattice model which neglects other degrees of freedom such as, translational or vibrational i.e. the molecular flexibility and considers only the orientational movement, is expected to help understand the important role of molecular biaxiality that plays in nematic - isotropic transition in bent-core systems as has been observed in the experimental investigation of Wiant $et~al.$ \cite{wiant}.
\section{ACKNOWLEDGMENT}
\indent N.G. acknowledges support through the Minor Research
Fellowship of University Grants Commission (PSW141/14-
15(ERO)); S.DG. acknowledges support through a research
grant obtained from Council of Scientific and Industrial
Research (03/1235/12/EMR-II). S.P. is thankful to University Grants Commission for support through a
fellowship.\\

\clearpage

\end{document}